\documentclass[10pt, draft, onecolumn, twoside]{IEEEtran}
\ifCLASSINFOpdf
\else
\fi
\usepackage{amsmath}
\usepackage{amsfonts}
\usepackage[dvipdfmx]{graphicx,xcolor}
\begin{document}
\title{Variable-Length Lossy Compression Allowing Positive Overflow and Excess Distortion Probabilities}

\author{Shota~Saito,~\IEEEmembership{Member,~IEEE,} Hideki~Yagi,~\IEEEmembership{Member,~IEEE,}
        and~Toshiyasu~Matsushima,~\IEEEmembership{Member,~IEEE}% <-this % stops a space
\thanks{%Manuscript received; revised.
This work was supported in part by JSPS KAKENHI Grant Numbers 
JP16K00195, %Yoshida
JP16K00417, %Maeda
JP16K06340, %Yagi
JP17K00316, %Ukita
JP17K06446, %Matsushima
and JP18K11585. %Saito
This paper was presented in part at the 2017 IEEE International Symposium on Information Theory (ISIT) \cite{Saito17} and the 2018 International Symposium on Information Theory and Its Applications (ISITA) \cite{Saito18}.}
\thanks{Shota Saito and Toshiyasu Matsushima are with the Department of Applied Mathematics, Waseda University, Tokyo 169-8555, Japan (e-mail: shota@aoni.waseda.jp; toshimat@waseda.jp).

Hideki Yagi is with the Department of Computer \& Network Engineering, The University of Electro-Communications, Tokyo 182-8585, Japan (e-mail: h.yagi@uec.ac.jp)}% <-this % stops a space
}

\maketitle

%150 - 250 words
\begin{abstract}
This paper investigates the problem of variable-length lossy source coding allowing a positive excess distortion probability and an overflow probability of codeword lengths. Novel one-shot achievability and converse bounds of the optimal rate are established by a new quantity based on the smooth max entropy (the smooth R\'enyi entropy of order zero). To derive the achievability bounds, we give an explicit code construction based on a distortion ball instead of using the random coding argument. The basic idea of the code construction is similar to the optimal code construction in the variable-length lossless source coding. Our achievability bounds are slightly different, depending on whether the encoder is stochastic or deterministic. One-shot results yield a general formula of the optimal rate for blocklength $n$. In addition, our general formula is applied to asymptotic analysis for a stationary memoryless source. As a result, we derive a single-letter characterization of the optimal rate by using the rate-distortion and rate-dispersion functions.
\end{abstract}


% \fi
\IEEEpeerreviewmaketitle

\newtheorem{theorem}{Theorem}
\newtheorem{condi}{Condition}
\newtheorem{defi}{Definition}
\newtheorem{lem}{Lemma}
\newtheorem{cor}{Corollary}
\newtheorem{proof}{Proof}
\newtheorem{rem}{Remark}
\newcommand{\argmax}{\mathop{\rm arg~max}\limits}
\newcommand{\argmin}{\mathop{\rm arg~min}\limits}

\section{Introduction} \label{INTRO}

\IEEEPARstart{T}{he} problem of source coding is one of the important research topics in Shannon theory.
Let $X^n = X_1 X_2 \ldots X_n$ be a source sequence taking a value in ${\cal X}^n$, where ${\cal X}$ is a source alphabet and ${\cal X}^n$ is the $n$-th Cartesian product of ${\cal X}$.
It is a recent active area of research to investigate the fundamental limits of the problem of source coding in the {\it non-asymptotic} regime, i.e., in the case where the blocklength $n$ is finite.

Regarding the non-asymptotic analysis for the problem of variable-length {\it lossless} source coding, Kontoyiannis and Verd\'u \cite{Kontoyiannis14} have studied the optimal rate.
In this study, the evaluation criterion of codeword lengths is the {\it overflow probability}, which is defined as the probability of codeword lengths per source symbol exceeding a certain threshold $R$.
More precisely, the overflow probability is defined as
\begin{align}
\Pr \left \{ \frac{1}{n} \ell(f_n (X^n)) >  R \right \},
\end{align}
where 
$
f_n : {\cal X}^n \rightarrow \{ 0,1 \}^{\star}  := \{\lambda, 0, 1, 00, \ldots \}\footnote{The notation $\{ 0,1 \}^{\star}$ denotes the set of all binary strings and the empty string $\lambda$.}
$
is an injective mapping called an encoder and the codeword lengths of $f_n (X^n)$ is denoted as $\ell(f_n (X^n))$.
The optimal rate at finite blocklength $n$ is characterized by the quantity $R^* (n, \delta)$: the lowest rate $R$ such that the overflow probability of the best code is not greater than $\delta \in [0, 1)$, i.e., 
\begin{align}
\min_{f_n} \Pr \left \{ \frac{1}{n} \ell(f_n (X^n)) >  R \right \} \leq \delta. \label{losslessoverflow}
\end{align}
In \cite{Kontoyiannis14}, the construction of the optimal encoder $f^{*}_{n}$ which achieves the minimum in (\ref{losslessoverflow}) is shown as follows.
Let $x^{n}_{i}$ be a source sequence which has the $i$-th largest probability, i.e., it holds that $P_{X^n} (x^{n}_{1}) \geq P_{X^n} (x^{n}_{2}) \geq P_{X^n} (x^{n}_{3}) \geq \cdots $.
Then, the encoder $f^{*}_{n}$ maps a source sequence $x^{n}_{1}, x^{n}_{2}, x^{n}_{3}, \ldots$ to the elements of $\{ 0,1 \}^{\star}$ in the lexicographic order, i.e.,
\begin{align}
f^{*}_{n} (x^{n}_{1}) & = \lambda,  \\
f^{*}_{n} (x^{n}_{2}) &=0, \\
f^{*}_{n} (x^{n}_{3}) &=1, \\
f^{*}_{n} (x^{n}_{4}) &=00, \\
f^{*}_{n} (x^{n}_{5}) &=01, \\
f^{*}_{n} (x^{n}_{6}) &=10, \\
f^{*}_{n} (x^{n}_{7}) &=11, \\
f^{*}_{n} (x^{n}_{8}) & =000, \\
& \vdots \notag
\end{align}
where $\lambda$ denotes the empty string.

Regarding the non-asymptotic analysis for the problem of variable-length {\it lossy} source coding, on the other hand, the code construction is more complicated because we have to consider a distortion between a source sequence and a reproduction sequence.
This fact raises the following question:
\begin{itemize}
\item[($\clubsuit$)] {\it Is it possible to extend the code construction in \cite{Kontoyiannis14} to the problem of variable-length lossy source coding?}
\end{itemize}

Motivated by the question ($\clubsuit$), we investigate the problem of variable-length lossy source coding under the criteria of the overflow probability of codeword lengths and the excess distortion probability.\footnote{Although the average distortion is a popular criterion of a distortion measure,  we treat the excess distortion probability because it is a natural way to look at lossy source coding problems at finite blocklength (see, e.g., \cite[Section 1.8.2]{Kostina13}).}
Thus, roughly speaking, we consider the following codes and the fundamental limit (the precise problem formulation is given in Sections \ref{OSC} and \ref{ASC}):
\begin{defi}
Let ${\cal X}$ be a source alphabet and ${\cal Y}$ be a reproduction alphabet.
Also, let ${\cal X}^n$ and ${\cal Y}^n$ be the $n$-th Cartesian product of ${\cal X}$ and ${\cal Y}$, respectively.
Let
$
d_n : {\cal X}^n \times {\cal Y}^n \rightarrow [0, +\infty)
$
be a distortion measure between a source sequence and a reproduction sequence.
Then, given $D, R\geq 0$ and $\epsilon, \delta \in [ 0,1)$, 
an encoder 
$
f_n : {\cal X}^n \rightarrow \{ 0,1 \}^{\star}
$
and a decoder 
$
g_n :  \{ 0,1 \}^{\star}  \rightarrow {\cal Y}^n
$ 
satisfying
\begin{align}
\Pr \left \{ \frac{1}{n} d_n (X^n, g_n (f_n (X^n))) > D \right \} &\leq \epsilon, \\
\Pr \left \{ \frac{1}{n} \ell(f_n (X^n)) > R \right \} &\leq \delta 
\end{align}
is called an $(n, D, R, \epsilon, \delta)$ code. 
The corresponding fundamental limit is 
\begin{align}
R^{*} & (n, D, \epsilon, \delta) := \inf \{ R :\mbox{$\exists$ an $(n, D, R, \epsilon, \delta)$ {\rm code }} \}. 
\end{align}
In particular, the optimal rate $R^{*} (n, D, \epsilon, \delta)$ at blocklength $n=1$ is abbreviated as 
$R^{*} (D, \epsilon, \delta) = R^{*} (1, D, \epsilon, \delta)$.
\end{defi}

\subsection{Contributions} \label{Contribution}
Our first contribution (and our main contribution) of this paper is to prove the one-shot bounds of the optimal rate $R^{*} (D, \epsilon, \delta)$ by a new quantity based on the smooth max entropy\footnote{
The smooth max entropy has first introduced by Renner and Wolf \cite{Renner}. The optimal rates for several problems have been characterized by the smooth max entropy
(e.g., \cite{Saito1}, \cite{Saito2}, \cite{Uyematsu1}, \cite{Uyematsu2}).
}(also known as the smooth R\'enyi entropy of order zero).
To derive the achievability bounds, we shall not use the random coding argument but give an explicit code construction. 
The code construction is similar to Feinstein's cookie-cutting argument \cite{Feinstein} and is based on a distortion $D$-ball centered at $y$:
\begin{align}
B_D (y) = \{ x \in {\cal X} : d (x, y) \leq D \} \label{IntroBD}
\end{align}
for a given $D \geq 0$, a distortion measure $d : {\cal X} \times {\cal Y} \rightarrow [0, +\infty)$, and a reproduction symbol $y \in {\cal Y}$.
As we show in Sec.\ \ref{PROOF}, this code construction gives the positive answer to the question ($\clubsuit$).
We derive the achievability bounds for both stochastic and deterministic encoders and show the difference between them.
On the other hand, the converse bounds are naturally derived from the definition of the new quantity based on the smooth max entropy.

Our second contribution of this paper is to give a general formula of the optimal rate $R^{*} (n, D, \epsilon, \delta)$.
The formula is easily derived from our one-shot achievability and converse bounds.
Our result indicates that the optimal rate $R^{*} (n, D, \epsilon, \delta)$ is the same regardless of the values of $\epsilon$ and $\delta$ if the sum of $\epsilon$ and $\delta$ is constant.
This result can be seen as the lossy version of the known result in the lossless source coding; see Remark \ref{e+d}.

Our third contribution of this paper is to derive a single-letter characterization of the optimal rate $R^{*} (n, D, \epsilon, \delta)$ for a stationary memoryless source.  
We apply our general theorem for a stationary memoryless source and characterize the optimal rate by using the rate-distortion and rate-dispersion functions.

\subsection{Related Works} \label{relatedwork}
As we mention in Remark \ref{fixed}, our problem setup is related to the problem of {\it fixed-length lossy source coding under the excess distortion probability}.
For this problem, the asymptotic analysis of the exponent of the excess distortion probability has been given by \cite{Ihara} and \cite{Marton}.
On the other hand, the non-asymptotic analysis of the minimum rate has been provided by \cite{Ingber} and \cite{Kostina12}.
When we prove a single-letter characterization of the optimal rate $R^{*} (n, D, \epsilon, \delta)$ for a stationary memoryless source, we utilize the result in \cite{Kostina12}.

For the problem of {\it variable-length lossy source coding under the excess distortion probability}, there are several criteria on codeword lengths: 
for example, the mean codeword lengths which is discussed in \cite{Kostina15};
the pointwise redundancy rate in \cite{Kontoyiannis00}, \cite{Kontoyiannis02};
the cumulant generating function of codeword lengths in \cite{Courtade}, \cite{Saito18ISIT};
the excess-code-length exponent in \cite{Weissman};
and the overflow probability in \cite{Nomura2}, \cite{Yagi}.
Among them,  Nomura and Yagi \cite{Nomura2} and Yagi and Nomura \cite{Yagi} adopt the same criteria in this paper: the overflow probability and the excess distortion probability.
However, the primary differences between these studies \cite{Nomura2}, \cite{Yagi} and our study are 
\begin{itemize}
\item[1)] we address the case where {\it both} the excess distortion probability and the overflow probability may be positive;
\item[2)] we analyze {\it both non-asymptotic and  asymptotic} cases, whereas the previous studies \cite{Nomura2}, \cite{Yagi} have investigated the asymptotic case;
\item[3)] we characterize the optimal rate by using the quantity related to the {\it entropy}, whereas previous studies \cite{Nomura2}, \cite{Yagi} have characterized it by using the quantity related to the {\it mutual information}.
\end{itemize}

It should be noted that the distortion $D$-ball in the previous studies and the distortion $D$-ball in our study are different. 
In the previous studies (e.g., \cite{Kontoyiannis00}, \cite{Kontoyiannis02}, \cite{Kostina12},  \cite{Kostina15}),
the distortion $D$-ball centered at $x$, i.e., the quantity
\begin{align}
\{ y \in {\cal Y} : d(x,y) \leq D \} \label{IntroBDP}
\end{align}
plays an important role.
On the other hand, as we have shown in Sec.\ \ref{Contribution}, our non-asymptotic analysis relies on the distortion $D$-ball centered at $y$ (i.e., (\ref{IntroBD})).
These differences mainly come from the way of the proof of the achievability result.
The previous studies have shown the achievability result by using the random coding argument.
In the random coding argument, the distortion $D$-ball centered at $x$ (i.e., (\ref{IntroBDP})) is a useful tool.
On the other hand, our study shows the achievability results by the explicit code construction.
In the code construction, the distortion $D$-ball centered at $y$ is a useful tool.

\subsection{Organization of the Paper}
The organization of this paper is as follows.
Section \ref{OSC} shows one-shot coding theorems:
we describe the problem setup in Sec.\ \ref{OSC_A},
define a new quantity based on the smooth max entropy in Sec.\ \ref{OSC_B},
and give one-shot coding theorems for stochastic and deterministic codes in Sec.\ \ref{OSC_C} and Sec.\ \ref{OSC_D}, respectively.
In Sec.\ \ref{PROOF}, proofs of the one-shot achievability bounds are described in detail.
Section \ref{ASC} shows general coding theorem for blocklength $n$:
we describe the problem formulation in Sec.\ \ref{ASC_A} and show a general formula of the optimal rate for blocklength $n$ in Sec.\ \ref{ASC_B}.
In Sec.\ \ref{ASC_C}, we apply our general formula to a stationary memoryless source and provide a single-letter characterization of the optimal rate.
Finally, Sec. \ref{Conc} concludes this paper.

\section{One-Shot Coding Theorem} \label{OSC}

\subsection{Problem Formulation} \label{OSC_A}
Let ${\cal X}$ be a source alphabet and ${\cal Y}$ be a reproduction alphabet, where both are finite sets.
Let $X$ be a random variable 
taking a value in ${\cal X}$ and $x$ be a realization of $X$.
The probability distribution of $X$ is denoted as $P_{X}$.
A distortion measure $d$ is defined as
$
d : {\cal X} \times {\cal Y} \rightarrow [0, +\infty).
$

The pair of an encoder and a decoder $(f, g)$ is defined as follows.
An encoder $f$  is defined as
$
f : {\cal X} \rightarrow \{ 0,1 \}^{\star},
$
where $\{ 0,1 \}^{\star}$ denotes the set of all binary strings 
and the empty string $\lambda$, i.e., 
\begin{align}
\{ 0,1 \}^{\star} := \{\lambda, 0, 1, 00, 01, 10, 11, 000, \ldots \}.
\end{align}
An encoder $f$ is possibly {\it stochastic} and produces a non-prefix code.
For $x \in {\cal X}$, the codeword lengths of $f(x)$ is denoted as $\ell(f(x))$.
A {\it deterministic} decoder $g$ is defined as
$
g :  \{ 0,1 \}^{\star}  \rightarrow {\cal Y}.
$
Variable-length lossy source coding {\it without} the prefix condition is discussed as in, for example, \cite{Courtade}, \cite{Kostina15}, and \cite{Nomura17}.

The performance criteria considered in this paper are the excess distortion and the overflow probabilities.

\begin{defi}
Given $D \geq 0$, 
the excess distortion probability for a code $(f, g)$ is defined as
\begin{align}
\Pr \{ d(X, g (f (X))) > D \}.
\end{align}
\end{defi}

\begin{defi}
Given $R \geq 0$, 
the overflow probability for a code $(f, g)$ is defined as
\begin{align}
\Pr \left \{ \ell(f (X)) > R \right \}.
\end{align}
\end{defi}

Using these criteria, we define a $(D, R, \epsilon, \delta)$ code.
\begin{defi}
Given $D, R\geq 0$ and $\epsilon, \delta \in [0,1)$, 
a code $(f, g)$ satisfying
\begin{align}
\Pr \{ d(X, g (f (X))) > D \} &\leq \epsilon \label{NP1}, \\
\Pr \left \{ \ell(f (X)) > R \right \} &\leq \delta  \label{NP2}
\end{align}
is called a $(D, R, \epsilon, \delta)$ code.
\end{defi}

The fundamental limits which we investigate are the following optimal rates
$R^{*} (D, \epsilon, \delta)$ and $\tilde{R} (D, \epsilon, \delta)$ for given $D$, $\epsilon$, and $\delta$.
\begin{defi}
Given $D\geq 0$ and $\epsilon, \delta \in  [0,1)$, 
\begin{align}
R^{*} (D, \epsilon, \delta) & := \inf \{ R : \mbox{$\exists$ {\rm a} $(D, R, \epsilon, \delta)$ {\rm code}} \}, \\
\tilde{R} (D, \epsilon, \delta) & := \inf \{ R : \mbox{$\exists$  {\rm a deterministic} $(D, R, \epsilon, \delta)$ {\rm code}} \}.
\end{align}
\end{defi}

\begin{rem} \label{fixed}
Consider the special case $\delta = 0$.
From any given $(D,R,\epsilon,0)$ code, we can construct a fixed-length code with the codeword length $\lfloor R \rfloor + 1$ and the excess distortion probability $\leq \epsilon$.
Thus, the special case $\delta = 0$ in our setup is closely related to the fixed-length source coding.
\end{rem}

\subsection{New Quantity based on the Smooth Max Entropy} \label{OSC_B}
The {\it smooth max entropy}, which is also called the {\it smooth R\'enyi entropy of order zero}, has first introduced by Renner and Wolf \cite{Renner}.
Later, Uyematsu \cite{Uyematsu1} has shown that the smooth max entropy can be defined 
in the following form.

\begin{defi}[\cite{Renner}, \cite{Uyematsu1}]
Given $\delta \in [0,1)$, the smooth max entropy $H^{\delta}(X)$ is defined as\footnote{Throughout this paper, all logarithms are of base 2 and $\exp \{ \cdot \}$ denotes $2^{(\cdot)}$.}
\begin{align}
H^{\delta}(X) := \min_{\substack{{\cal Z} \subset {\cal X} : \\
\Pr \{ X \in {\cal Z} \} \geq 1-\delta}} \log |{\cal Z}|, \label{SME}
\end{align}
where $|\cdot|$ represents the cardinality of the set.
\end{defi}

One of the useful properties of the smooth max entropy, which is used in the proof of the achievability result in our main theorem, is the {\it Schur concavity}.
To state the definition of a Schur concave function, we first review the notion of {\it majorization}.

\begin{defi}
Let $\mathbb{R}_{+}$ be the set of non-negative real numbers 
and $\mathbb{R}^{m}_{+}$ be the $m$-th Cartesian product of $\mathbb{R}_{+}$, where $m$ is a positive integer.
Suppose that ${\bf a} = (a_1, \ldots,$ $a_m) \in \mathbb{R}^{m}_{+}$ and ${\bf b} = (b_1, \ldots, b_m) \in \mathbb{R}^{m}_{+}$ satisfy
\begin{align}
a_i \geq a_{i+1}, \quad b_i \geq b_{i+1} \quad (i=1,2, \ldots, m-1).
\end{align}
If ${\bf a} \in \mathbb{R}^{m}_{+}$ and ${\bf b} \in \mathbb{R}^{m}_{+}$ satisfy, for $k=1, \ldots, m-1$, 
\begin{align}
\sum_{i=1}^{k} a_i \leq \sum_{i=1}^{k} b_i  \quad  {\rm and } \quad \sum_{i=1}^{m} a_i = \sum_{i=1}^{m} b_i ,
\end{align}
then we say that ${\bf b}$ {\it majorizes} ${\bf a}$ (it is denoted as ${\bf a} \prec {\bf b}$ in this paper).
\end{defi}

Schur concave functions are defined as follows.
\begin{defi}
We say that a function $h(\cdot): \mathbb{R}^{m}_{+} \rightarrow \mathbb{R}$ is a {\it Schur concave} function
if $h({\bf b}) \leq h({\bf a})$ for any ${\bf a}, {\bf b} \in \mathbb{R}^{m}_{+}$ satisfying ${\bf a} \prec {\bf b}$.
\end{defi}

From the definition of the smooth max entropy and Schur concave functions, it is easy to see that the smooth max entropy is a Schur concave function\footnote{
In \cite{Koga}, 
by using the notion of majorization, it is shown that the smooth R\'enyi entropy of order $\alpha$ is a Schur concave function for $0 \leq \alpha < 1$ and a Schur convex function for $\alpha > 1$.}.

Next, using the smooth max entropy, we introduce a new quantity, which plays an important role in producing our main results.

\begin{defi}
Given $D \geq 0$ and $\epsilon, \delta \in [0, 1)$, $G_{D, \epsilon}^{\delta}(X)$ is defined as
\begin{align}
G_{D, \epsilon}^{\delta}(X) 
&:= 
\inf_{\substack{P_{Y|X} : \\
\Pr \{ d(X, Y) > D \} \leq \epsilon}} H^{\delta}(Y) \label{G} \\
&= \inf_{\substack{P_{Y|X} : \\
\Pr \{ d(X, Y) > D \} \leq \epsilon}} ~~
\min_{\substack{{\cal W} \subset {\cal Y} : \\
\Pr \{ Y \in {\cal W} \} \geq 1-\delta}} \log |{\cal W}|, 
\end{align}
where $P_{Y|X}$ denotes a conditional probability distribution of $Y$ given $X$.
\end{defi}

\begin{rem}
For a given  $D \geq 0$ and $\epsilon \in [0,1)$, suppose that
\begin{align}
\Pr \left \{ \min_{y \in {\cal Y}} d(X, y) >D \right \} >\epsilon. \label{inf}
\end{align}
Then, there are no codes whose excess distortion probability is less than or equal to $\epsilon$.
Conversely, if such codes do not exist for given $D$ and $\epsilon$, (\ref{inf}) holds.
In this case, we define $R^{*} (D, \epsilon, \delta) = +\infty$ and $\tilde{R} (D, \epsilon, \delta) = +\infty$.
Further, if (\ref{inf}) holds, we also define $G_{D, \epsilon}^{\delta}(X) = +\infty$
because there is no conditional probability distribution $P_{Y|X}$ on ${\cal Y}$ satisfying
$\Pr \{ d(X, Y) > D \} \leq \epsilon$.
\end{rem}

\begin{rem}
As shown in Theorems \ref{Th_oneshot_sc}, \ref{Th_oneshot_dt}, and \ref{Th_as_g}, $G_{D, \epsilon}^{\delta}(X)$ is a crucial quantity in characterizing the optimal rate.
Hence, it is worth mentioning the property of $G_{D, \epsilon}^{\delta}(X)$, which indicates that the value of  $G_{D, \epsilon}^{\delta}(X)$ only depends on the sum of $\epsilon$ and $\delta$.
We shall use this property to discuss our result (see Remark \ref{e+d}).
\begin{lem} \label{PropertyG}
Fix $\epsilon, \delta \in [0, 1)$ arbitrarily.
Then, for any $\epsilon', \delta' \in [0, 1)$ such that $\epsilon' + \delta' = \epsilon + \delta$, we have
\begin{align}
G_{D, \epsilon}^{\delta}(X) = G_{D, \epsilon'}^{\delta'}(X). 
\end{align}
\end{lem}
\begin{IEEEproof}
See Appendix \ref{PropG}.
\end{IEEEproof}
\end{rem}

\subsection{One-Shot Coding Theorem for Stochastic Codes} \label{OSC_C}
The next lemma shows the achievability result on $R$ of a $(D, R, \epsilon, \delta)$ code.

\begin{lem} \label{Lem_oneshot_ac}
Assume that $G_{D, \epsilon}^{\delta}(X) < + \infty.$
For any $D \geq 0$ and $\epsilon, \delta \in [0, 1)$,
there exists a $(D, R, \epsilon, \delta)$ code such that
\begin{align}
R = \lfloor G_{D, \epsilon}^{\delta}(X) \rfloor. \label{TH5}
\end{align}
\end{lem}

\begin{IEEEproof}
See Sec.\ \ref{Proof_Lem_oneshot_ac}.
\end{IEEEproof}

The next lemma shows the converse bound on $R$ of a $(D, R, \epsilon, \delta)$ code.

\begin{lem} \label{Lem_oneshot_con}
Assume that $G_{D, \epsilon}^{\delta}(X) < + \infty.$
For any $D \geq 0$ and $\epsilon, \delta\in [0, 1)$,
any  $(D, R, \epsilon, \delta)$ code satisfies
\begin{align}
R > G_{D, \epsilon}^{\delta}(X) -1. \label{NTH1}
\end{align}
\end{lem}

\begin{IEEEproof}
For any $(D, R, \epsilon, \delta)$ code $(\overline{f}, \overline{g})$,  
set $\overline{Y}:= \overline{g}(\overline{f} (X))$.
To prove the converse bound, it is sufficient to consider the case where the decoder $\overline{g}$ is an injective mapping\footnote{Suppose that $\overline{g}$ is not an injective mapping. That is, $\overline{g}(w) = \overline{g}(w')$ holds for $w \neq w'$, $w = \overline{f}(x)$ and $w' = \overline{f}(x')$.
Further, suppose that $\ell(w) \leq \ell(w')$ holds.
In this case, by adjusting the encoder as $w = \tilde{f}(x)$, $w = \tilde{f}(x')$, we can make the codeword length shorter without affecting the excess distortion probability. This modification makes the decoder injective.}.
The definition of a $(D, R, \epsilon, \delta)$ code gives
\begin{align}
\Pr \left \{ \ell(\overline{f} (X)) > R \right \} & \leq \delta, \label{NPCF} \\
\Pr \{ d(X, \overline{Y}) > D \} & \leq \epsilon. \label{NPCS} 
\end{align}
Let $T$ be defined as
\begin{align}
T :=  \left \{ \overline{g} (\overline{f}(x)) \in {\cal Y} : x  ~ {\rm satisfies } ~ \ell(\overline{f}(x)) > R  \right \}.
\end{align}
Then, (\ref{NPCF}) is rewritten as
\begin{align}
\Pr \left \{ \overline{Y} \in T \right \} \leq \delta.  
\end{align}
Hence, 
\begin{align}
\Pr \left \{ \overline{Y} \in T^{c} \right \} \geq 1 - \delta, \label{NPC3}
\end{align}
where the superscript ``$c$'' represents the complement. 
From (\ref{NPC3}) and the definition of the smooth max entropy,
we have
\begin{align}
H^{\delta}(\overline{Y}) \leq \log | T^c | \label{NPC4}.
\end{align}

On the other hand, since $\ell(\overline{g}^{-1}(y)) \leq  \lfloor R \rfloor$ for $ y \in T^c$, 
\begin{align}
\hspace{-3mm} | T^c |
\leq
1+2+\cdots+2^{\lfloor R \rfloor} 
= 2^{\lfloor R \rfloor + 1} - 1 
< 2^{R+1}. \label{NPC5}
\end{align}
Combining (\ref{NPC4}) and (\ref{NPC5}) yields
\begin{align}
H^{\delta}(\overline{Y}) < R+1. 
\end{align}
Thus, from (\ref{NPCS}), we have
\begin{align}
G_{D, \epsilon}^{\delta}(X)  < R+1.
\end{align}
Hence, we obtain (\ref{NTH1}).
\end{IEEEproof}

Combining Lemmas \ref{Lem_oneshot_ac} and \ref{Lem_oneshot_con}, 
we can immediately obtain the following result on $R^{*} (D, \epsilon, \delta)$.

\begin{theorem} \label{Th_oneshot_sc}
Assume that $G_{D, \epsilon}^{\delta}(X) < + \infty.$
For any $D \geq 0$ and $\epsilon, \delta \in [0, 1)$,
it holds that
\begin{align}
G_{D, \epsilon}^{\delta}(X)  - 1 < R^{*}(D, \epsilon, \delta) \leq \lfloor G_{D, \epsilon}^{\delta}(X) \rfloor. \label{NEP}
\end{align}
\end{theorem}

By Theorem \ref{Th_oneshot_sc}, the optimal rate $R^{*}(D, \epsilon, \delta)$ can be specified within one bit in the interval not greater than $G_{D, \epsilon}^{\delta}(X)$, 
regardless of the values $D$, $\epsilon$, and $\delta$.
This result is mainly due to an explicit construction of good codes, rather than the random coding argument, 
given in Sec.\ \ref{Proof_Lem_oneshot_ac}.

\subsection{One-Shot Coding Theorem for Deterministic Codes} \label{OSC_D}

The next lemma shows the achievability result on $R$ of a deterministic $(D, R, \epsilon, \delta)$ code.

\begin{lem} \label{Lem_oneshot_ac_dt}
Assume that $G_{D, \epsilon}^{\delta}(X) < + \infty.$
For any $D \geq 0$ and $\epsilon, \delta \in [0, 1)$,
there exists a deterministic $(D, R, \epsilon, \delta)$ code such that
\begin{align}
R = \left \lfloor G_{D, \epsilon}^{\delta}(X) + \frac{2 \log e}{2^{G_{D, \epsilon}^{\delta}(X) }} \right \rfloor. \label{TH4}
\end{align}
\end{lem}

\begin{IEEEproof}
See Sec.\ \ref{Proof_Lem_oneshot_ac_dt}. 
\end{IEEEproof}

From Lemma \ref{Lem_oneshot_ac_dt} and the fact that $R^{*} (D, \epsilon, \delta) \leq \tilde{R} (D, \epsilon, \delta)$, 
the following result on $\tilde{R} (D, \epsilon, \delta)$ is obtained.

\begin{theorem} \label{Th_oneshot_dt}
Assume that $G_{D, \epsilon}^{\delta}(X) < + \infty.$
For any $D \geq 0$ and $\epsilon, \delta \in [0, 1)$,
it holds that
\begin{align}
G_{D, \epsilon}^{\delta}(X)  - 1  & < \tilde{R} (D, \epsilon, \delta)  \notag \\
& \leq \left \lfloor G_{D, \epsilon}^{\delta}(X) + \frac{2 \log e}{2^{G_{D, \epsilon}^{\delta}(X) }} \right \rfloor. \label{NEP}
\end{align}
\end{theorem}

By Theorem \ref{Th_oneshot_dt}, $\tilde{R} (D, \epsilon, \delta)$ can be specified in the interval within four bits, which is slightly weaker than the result for stochastic codes.

\section{Proofs of One-Shot Achievability Results} \label{PROOF}
\subsection{Outline of the Code Construction}
In the proof of the achievability result, we give an explicit code construction.
Before we formally describe it, we illustrate the gist of the construction.  

First, we define $B_D (y)$, the distortion $D$-ball centered at $y$, as (\ref{BD}).
Next, we label reproduction symbols as $y_1, y_2, \ldots$ according to the probability of the distortion $D$-ball; see (\ref{Y1}) and (\ref{Yi}).
Then, we modify the distortion $D$-ball and define the ball $A_D (y_i)$ so that each ball is disjoint and $\Pr \{ X \in A_{D} (y_1) \} \geq\Pr \{ X \in A_{D} (y_2) \} \geq \cdots$ holds; see (\ref{ad1})--(\ref{Kakuritsu}).
Then, by using the probability $\Pr \{ X \in A_{D} (y_i) \}$, we define the integer $k^*$ satisfying (\ref{ks2}) and (\ref{ks1}).
Now, roughly speaking, we construct the code by the following strategy:
\begin{itemize}
\item For a source symbol $x \in A_{D} (y_i)$ ($i=1, \ldots, k^*$) (i.e., a source symbol such that the probability $\Pr \{ X \in A_{D} (y_i) \}$ is large), we encode it not to exceed the distortion level $D$.
Further, for a source symbol $x \in A_{D} (y_i)$ ($i=1, \ldots, k^*$) such that the probability $\Pr \{ X \in A_{D} (y_i) \}$ is larger, we encode it to have shorter codeword\footnote{Strictly speaking, we encode a source symbol $x \in A_{D} (y_{k^*})$ differently whether we use stochastic encoders or deterministic enoders. If we use stochastic encoders to apply the randomized mapping to $x \in A_{D} (y_{k^*})$, 
\begin{itemize}
\item we can make the excess distortion probability exactly $\epsilon$, 
\item we can make the overflow probability exceeding the rate $\lfloor G_{D, \epsilon}^{\delta}(X) \rfloor$ less than or equal to $\delta$.
\end{itemize}
On the other hand, if we use deterministic encoders, the excess distortion probability is less than or equal to $\epsilon$, not exactly $\epsilon$.
In this case, to make the overflow probability smaller than $\delta$, we must set the rate
$
\left \lfloor G_{D, \epsilon}^{\delta}(X) + 2 \log e / 2^{G_{D, \epsilon}^{\delta}(X)} \right \rfloor,
$
which is slightly larger than $\lfloor G_{D, \epsilon}^{\delta}(X) \rfloor$.
}. Note that the basic idea is similar to the optimal code construction in the variable-length lossless source coding in Sec.\ \ref{INTRO}; for a source symbol $x$ such that the probability $P_X (x)$ is larger, we encode it to have shorter codeword.
\item For a source symbol $x \in A_{D} (y_i)$ ($i=k^* +1, \ldots$) (i.e., a source symbol such that the probability $\Pr \{ X \in A_{D} (y_i) \}$ is small), we encode it to exceed the distortion level $D$ and to have shortest codeword (i.e., the empty string $\lambda$).
\end{itemize}

Now, we formally explain the code construction in the following subsections.

\subsection{Proof of Lemma \ref{Lem_oneshot_ac}} \label{Proof_Lem_oneshot_ac}
First, some notations are defined before the construction of the encoder and the decoder is described.
\begin{itemize}
\item For $y \in {\cal Y}$ and $D \geq 0$, the distortion $D$-ball centered at $y$ is defined as
\begin{align}
B_D (y) := \{ x \in {\cal X} : d(x,y) \leq D \}. \label{BD}
\end{align}

\item We define  $y_i $ ($i = 1, 2, \cdots$) by the following procedure\footnote{
In this paper, we assume that ${\cal X}$ and ${\cal Y}$ are finite sets.
However, we can assume countably infinite ${\cal X}$ and ${\cal Y}$
if this operation is admitted for countably infinite ${\cal X}$ and ${\cal Y}$.
}.
Let $y_1$ be defined as
\begin{align}
y_1 := \argmax_{y \in {\cal Y}} \Pr \{ X \in B_{D} (y) \}, \label{Y1}
\end{align}
and for $i = 2, 3, \cdots$, let $y_i $ be defined as
\begin{align}
y_i := \argmax_{y \in {\cal Y}} \Pr \left \{ X \in B_{D} (y) \setminus \bigcup_{j=1}^{i-1} B_{D} (y_j) \right \}. \label{Yi}
\end{align}

\item For $i = 1, 2, \cdots$, we define $A_D(y_i)$ by 
\begin{align}
A_D(y_1) &:= B_{D} (y_1), \label{ad1} \\
A_D(y_i) &:= B_{D} (y_i) \setminus \bigcup_{j=1}^{i-1} B_{D} (y_j) \quad (\forall i \geq 2) \label{adi}
\end{align}
From the definition, we have
\begin{align}
& \bigcup_{j=1}^{i} A_{D} (y_j) = \bigcup_{j=1}^{i} B_{D} (y_j) \quad (\forall i \geq 1), \label{A1} \\
& A_{D} (y_i) \cap A_{D} (y_j) =  \emptyset \quad (\forall i \neq j), \label{A2} \\
& \hspace{-10mm} \Pr \{ X \in A_{D} (y_1) \} \geq\Pr \{ X \in A_{D} (y_2) \} \geq \cdots. \label{Kakuritsu}
\end{align}

\item If $\epsilon + \delta < 1$, let $i^{*} \geq 1$ be the integer satisfying 
\begin{align}
\sum_{i=1}^{i^{*}-1} \Pr \{ X \in A_{D} (y_i) \} &< 1- \epsilon - \delta, \label{ih2} \\
\sum_{i=1}^{i^{*}} \Pr \{ X \in A_{D} (y_i) \} & \geq 1- \epsilon - \delta. \label{ih1}
\end{align}
If $\epsilon + \delta \geq 1$, we define $i^* =1$.

\item Let $k^{*} \geq 1$ be the integer satisfying 
\begin{align}
\sum_{i=1}^{k^{*}-1} \Pr \{ X \in A_{D} (y_i) \} & < 1-\epsilon, \label{ks2}
\\
\sum_{i=1}^{k^{*}} \Pr \{ X \in A_{D} (y_i) \} & \geq 1-\epsilon. \label{ks1}  
\end{align}
From this definition, it holds that $k^{*} \geq i^{*}$.

\item Let $\alpha$ and $\beta$ be defined as
\begin{align}
\alpha := \sum_{i=1}^{k^{*}-1} \Pr \{ X \in A_{D} (y_i) \}
\end{align}
and
\begin{align}
\beta := 1-\epsilon-\alpha.
\end{align}

\item Let $w_{i}$ be the $i$-th binary string in $\{ 0,1 \}^{\star}$ in the increasing order of the length and ties are arbitrarily broken.
For example, $w_1 = \lambda, w_2 = 0, w_3 = 1, w_4 = 00, w_5 = 01,$ etc.
\end{itemize}

We construct the following encoder 
$
\hat{f} : {\cal X} \rightarrow \{ 0,1 \}^{\star}
$
and decoder
$
\hat{g} :  \{ 0,1 \}^{\star}  \rightarrow {\cal Y}.
$

\medskip

\noindent
{\bf [Encoder]}
\begin{itemize}
\item[$ 1)$] For $x \in A_{D} (y_i)$ ($i=1, \ldots, k^* -1$), set 
$
\hat{f}(x) = w_i.
$

\item[$2)$] For $x \in A_{D} (y_{k^*})$, set\footnote{
Note that we have $\Pr \{ X \in A_{D} (y_{k^*}) \} \geq \beta$ from (\ref{ks1}).
}
\begin{align}
\hspace{-12mm} \hat{f}(x) = \begin{cases}
    w_{k^*} &  {\rm with~ prob.} ~ \frac{\beta}{ \Pr \{ X \in A_{D} (y_{k^*}) \} }, \\
    w_1&  {\rm with~ prob.} ~ 1 - \frac{\beta}{ \Pr \{ X \in A_{D} (y_{k^*}) \} }.
  \end{cases}
\label{se}
\end{align}

\item[$3)$] For $x \notin  \bigcup_{i=1}^{k^*} A_{D} (y_i)$, set 
$
\hat{f}(x) = w_1.
$
\end{itemize}

\noindent
{\bf [Decoder]}
For $i=1, \ldots, k^* $, set 
$
\hat{g}(w_i) = y_i .
$
\medskip

Now, we evaluate the excess distortion probability.
We have
$
d(x, \hat{g}(\hat{f}(x))) \leq D
$
for $x \in A_{D} (y_i)$ ($i=1, \ldots, k^{*}-1$) since $\hat{g}(\hat{f}(x))=y_i$.
Furthermore, we have $d(x, \hat{g}(\hat{f}(x))) \leq D$ with probability $\beta/\Pr \{ X \in A_{D} (y_{k^*}) \}$ for
$x \in A_{D} (y_{k^*})$. Thus,
\begin{align}
\Pr & \{ d(X, \hat{g} (\hat{f} (X))) \leq D \} \notag \\
&= \sum^{k^{*}-1}_{i=1} \Pr \{ X \in A_{D} (y_i) \} + \Pr \{ \hat{f}(X) = w_{k^*}, X \in A_{D} (y_{k^*}) \} \\
&=\alpha + \beta \\
&= 1 - \epsilon. 
\end{align}
Therefore, we have
\begin{align}
\Pr \{ d(X, \hat{g} (\hat{f} (X))) > D \} = \epsilon. \label{hc}
\end{align}

Next, we evaluate the overflow probability.
From the construction of the encoder, it is easily verified that
$\ell(w_i) = \lfloor \log i \rfloor$ ($i= 1, \ldots, k^*$).
Hence, setting $R = \lfloor \log i^* \rfloor$, we have
\begin{align}
\Pr & \left \{ \ell(\hat{f}(X)) > R \right \} \notag \\
& \leq \sum^{k^{*}}_{i=i^{*}+1} \Pr \{ \hat{f}(X) = w_i \} \\
& = \sum^{k^{*}-1}_{i=i^{*}+1} \Pr \{ X \in A_{D} (y_i) \} + \Pr \{ \hat{f}(X) = w_{k^*}, X \in A_{D} (y_{k^*}) \} \\
& = \sum^{k^{*}-1}_{i=1} \Pr \{ X \in A_{D} (y_i) \} - \sum^{i^{*}}_{i=1} \Pr \{ X \in A_{D} (y_i) \} +\beta \\
& \leq \alpha - (1 - \epsilon - \delta) + \beta 
 = \delta,
\end{align}
where the last inequality is due to the definition of $\alpha$ and (\ref{ih1}) and the last equality is due to the definition of $\beta$.

Therefore, the code $(\hat{f}, \hat{g})$ is a $(D, R, \epsilon, \delta)$ code with $R = \lfloor \log i^* \rfloor$.
To complete the proof, we shall show 
\begin{align}
\log i^* = G_{D, \epsilon}^{\delta}(X). \label{iG}
\end{align}
To this end, we define $\hat{Y} := \hat{g}(\hat{f}(X))$.
Then, we have the next lemma.
\begin{lem} \label{HGLemma}
For any $D \geq 0$ and $\epsilon, \delta \in [0,1)$, we have
\begin{align}
H^{\delta}(\hat{Y}) = G_{D, \epsilon}^{\delta}(X) \label{HG}
\end{align}
\end{lem}
\begin{IEEEproof}
See Appendix \ref{Append1}.
\end{IEEEproof}
In view of Lemma \ref{HGLemma}, if we show
\begin{align}
\log i^* = H^{\delta}(\hat{Y}), \label{logi H}
\end{align}
we obtain the desired equation (\ref{iG}).
Therefore, we shall show (\ref{logi H}) in the rest of the proof.

First, notice that
\begin{align}
P_{\hat{Y}}(y_1) 
&\overset{(a)}{=} \Pr \{ X \in A_{D} (y_1) \}  + \Pr \{ X \in \bigcup_{i \geq k^{*}+1} A_{D} (y_i) \} + \Pr \{ \hat{f}(X) =w_1, X \in A_{D} (y_{k^*}) \} \\
& \overset{(b)}{=} \Pr \{ X \in A_{D} (y_1) \} + \Pr \{ d(X, \hat{g} (\hat{f} (X))) > D \} \\
& \overset{(c)}{=}  \Pr \{ X \in A_{D} (y_1) \} + \epsilon, \label{y11} \\
P_{\hat{Y}}(y_i) &= \Pr \{ X \in A_{D} (y_i) \}
\quad (i = 2, \ldots, k^* - 1) \label{y12},
\end{align}
where $(a)$ and $(b)$ follow from the definition of the encoder and the decoder and $(c)$ is due to (\ref{hc}).
Then\footnote{If $i^* = k^*$, the equality in (\ref{notice}) does not hold. However, $\sum^{i^{*}}_{i=1} P_{\hat{Y}} (y_i)  \geq 1-\delta$ is true since $\sum^{i^{*}}_{i=1} P_{\hat{Y}} (y_i) = 1$.}, 
\begin{align}
&\sum^{i^{*}-1}_{i=1} P_{\hat{Y}}(y_i) 
= \sum^{i^{*}-1}_{i=1} \Pr \{ X \in A_{D} (y_i) \} + \epsilon
< 1 - \delta, \\
&\sum^{i^{*}}_{i=1} P_{\hat{Y}}(y_i) 
=\sum^{i^{*}}_{i=1} \Pr \{ X \in A_{D} (y_i) \} + \epsilon
\geq 1 - \delta, \label{notice} \\
&P_{\hat{Y}} (y_1) \geq P_{\hat{Y}}(y_2) \geq \cdots \geq P_{\hat{Y}} (y_{k^*}), 
\end{align}
which imply that $\log i^* = H^{\delta}(\hat{Y})$.

\subsection{Proof of Lemma \ref{Lem_oneshot_ac_dt}} \label{Proof_Lem_oneshot_ac_dt}
First, some notations are defined.
\begin{itemize}
\item Let $k^{*} \geq 1$ be the integer satisfying (\ref{ks2}) and (\ref{ks1}).

\item Define $\gamma$ as 
\begin{align}
\gamma := 1- \sum_{i=1}^{k^{*}} \Pr \{ X \in A_{D} (y_i) \},
\end{align}
where $\{ y_1, y_2, \ldots \}$ and $A_{D} (y_i)$ are defined as in Sec.\ \ref{Proof_Lem_oneshot_ac}.
Then, it holds that $\gamma \leq \epsilon$.

\item Let $j^* \geq 1$ be the integer satisfying
\begin{align}
\sum_{i=1}^{j^{*}-1} \Pr \{ X \in A_{D} (y_i) \} & < 1-\gamma - \delta, \label{js2} 
\\
\sum_{i=1}^{j^{*}} \Pr \{ X \in A_{D} (y_i) \} & \geq 1- \gamma - \delta. \label{js1} 
\end{align}
\end{itemize}

We construct the following {\it deterministic} encoder 
$
\hat{f}_{\rm det} : {\cal X} \rightarrow \{ 0,1 \}^{\star}
$
and decoder
$
\hat{g}_{\rm det} :  \{ 0,1 \}^{\star}  \rightarrow {\cal Y}.
$

\noindent
{\bf [Encoder]}
\begin{itemize}
\item[$ 1)$] For $x \in A_{D} (y_i)$ ($i=1, \ldots, k^* $), set 
$
\hat{f}_{\rm det} (x) = w_i.
$

\item[$2)$] For $x \notin  \bigcup_{i=1}^{k^*} A_{D} (y_i)$, set 
$
\hat{f}_{\rm det} (x) = w_1.
$
\end{itemize}

\noindent
{\bf [Decoder]}
For $i=1, \ldots, k^* $, set 
$
\hat{g}_{\rm det}(w_i) = y_i
$
\medskip

Now, we evaluate the excess distortion probability.
From the definition of the encoder and the decoder, we have
\begin{align}
\Pr \{ d(X, \hat{g}_{\rm det} (\hat{f}_{\rm det} (X))) \leq D \} 
& = \sum^{k^{*}}_{i=1} \Pr \{ X \in A_{D} (y_i) \}  \\
& \ge 1-\epsilon.
\end{align}
Therefore, we have
$
\Pr \{ d(X, \hat{g}_{\rm det} (\hat{f}_{\rm det}(X))) > D \} \leq \epsilon.
$

Next, we evaluate the overflow probability.
From the definition of the encoder, we have
\begin{align}
& \Pr \{ \hat{f}_{\rm det}(X) = w_1 \} = \Pr \{X \in A_{D} (y_1) \} + \gamma,\\
& \Pr \{ \hat{f}_{\rm det}(X) = w_i \} = \Pr \{X \in A_{D} (y_i) \} \quad (i=2, \ldots, k^*).
\end{align}
Setting $R = \lfloor \log \min (j^* , k^* ) \rfloor$, it holds that\footnote{
If $R=\lfloor \log k^* \rfloor$, then $\Pr \{\ell (\hat{f}_{\rm det}(X)) > R \} =0$.}
\begin{align}
\Pr & \left \{ \ell(\hat{f}_{\rm det}(X)) > R \right \}  \notag \\
& \leq 1 - \sum^{j^{*}}_{i=1} \Pr \{ \hat{f}_{\rm det}(X) = w_i \} \\
&  = 1 - \left ( \sum^{j^{*}}_{i=1} \Pr \{ X \in A_{D} (y_i) \} + \gamma \right ) \\
& \leq 1- ( (1-\gamma - \delta) + \gamma) = \delta,
\end{align}
where
the last inequality is due to (\ref{js1}).

Therefore, the code $(\hat{f}_{\rm det}, \hat{g}_{\rm det})$ is a deterministic $(D, R, \epsilon, \delta)$ code with $R = \lfloor \log \min (j^* , k^* ) \rfloor$.
\medskip

Let $i^*$ be the integer satisfying (\ref{ih2}) and (\ref{ih1}).
Then, from the proof of Lemma 1, it holds that (see (\ref{iG}))
\begin{align}
\log i^* = G_{D, \epsilon}^{\delta}(X). \label{Ketsuron1}
\end{align}
Since $\gamma \leq \epsilon$, it is easily verified that $i^* \leq j^*$ and $i^* \leq k^*$, meaning that $i^* \leq \min(j^*, k^*)$.
If $i^* = \min(j^*, k^*)$, $\min(j^*, k^*) \leq i^* + 2$ obviously holds.
If $i^* < \min(j^*, k^*)$, we can show (Appendix \ref{Append2})
\begin{align}
\min(j^*, k^*) \leq i^* + 2. \label{IJK}
\end{align}
Therefore, we have
\begin{align}
\log \min(j^*, k^*) \leq \log (i^* +2) \leq \log i^* + \frac{2 \log e}{i^*},  \label{Ketsuron2}
\end{align}
where the rightmost inequality is due to Taylor's expansion.
Finally, combining (\ref{Ketsuron1}) and (\ref{Ketsuron2}), we conclude that (\ref{TH4}) holds.

\section{General Coding Theorem} \label{ASC}
\subsection{Problem Formulation} \label{ASC_A}
Let ${\cal X}^n$ and ${\cal Y}^n$ be the $n$-th Cartesian product of ${\cal X}$ and ${\cal Y}$, respectively.
Let $X^n$ be a random variable 
taking a value in ${\cal X}^n$ and $x^n$ be a realization of $X^n$.
The probability distribution of $X^n$ is denoted as $P_{X^n}$.
A distortion measure $d_n$ is defined as
$
d_n : {\cal X}^n \times {\cal Y}^n \rightarrow [0, +\infty).
$
An encoder 
$
f_n : {\cal X}^n \rightarrow \{ 0,1 \}^{\star}
$
is possibly stochastic and produces a non-prefix code.
A decoder 
$
g_n :  \{ 0,1 \}^{\star}  \rightarrow {\cal Y}^n
$
is deterministic.

We define an $(n, D, R, \epsilon, \delta)$ code as follows.
\begin{defi}
Given $D, R\geq 0$ and $\epsilon, \delta \in [ 0,1)$, 
a code $(f_n, g_n)$ satisfying
\begin{align}
\Pr \left \{ \frac{1}{n} d_n (X^n, g_n (f_n (X^n))) > D \right \} &\leq \epsilon, \label{cd1} \\
\Pr \left \{ \frac{1}{n} \ell(f_n (X^n)) > R \right \} &\leq \delta \label{cd2}
\end{align}
is called an $(n, D, R, \epsilon, \delta)$ code.
\end{defi}

The fundamental limits which we investigate are the following optimal rates:
\begin{align}
R^{*} & (n, D, \epsilon, \delta) := \inf \{ R :\mbox{$\exists$ an $(n, D, R, \epsilon, \delta)$ {\rm code }} \}. \\
\tilde{R} & (n, D, \epsilon, \delta) := \inf \{ R : \mbox{$\exists$  {\rm a deterministic} $(n, D, R, \epsilon, \delta)$ {\rm code}} \}.
\end{align}

\subsection{General Coding Theorem for Stochastic Codes and Deterministic Codes} \label{ASC_B}
The same discussion which is used to prove Theorems \ref{Th_oneshot_sc} and \ref{Th_oneshot_dt} establishes the next result on $R^{*} (n, D, \epsilon, \delta)$ and $\tilde{R} (n, D, \epsilon, \delta)$.

\begin{theorem} \label{Th_as_g}
For any $D \geq 0$ and $\epsilon, \delta \in [0, 1)$, we have
\begin{align} 
R^{*} (n, D, \epsilon, \delta)  = \tilde{R} (n, D, \epsilon, \delta) =  \frac{1}{n} G_{D, \epsilon}^{\delta}(X^n) + O \left( \frac{1}{n} \right ),
\end{align}
where 
\begin{align}
G_{D, \epsilon}^{\delta}(X^n) 
&:= 
\inf_{\substack{P_{Y^n |X^n } : \\
\Pr \{ d_n (X^n, Y^n) > nD \} \leq \epsilon}} H^{\delta}(Y^n).
\end{align}
\end{theorem}

\begin{rem}
In the one-shot coding regime, the results of the optimal rate between stochastic encoders and deterministic encoders are different as shown in Theorems \ref{Th_oneshot_sc} and \ref{Th_oneshot_dt}.
When we consider the setting of blocklength $n$, however, the restriction to only deterministic encoders does not change the result.
This is because the additional term 
$
2 \log e / n 2^{G_{D, \epsilon}^{\delta}(X^n) }
$
for deterministic codes is $O(1/n)$.
\end{rem}

\begin{rem} \label{e+d}
By combining Lemma \ref{PropertyG} and Theorem \ref{Th_as_g}, we see that the optimal rate $R^{*} (n, D, \epsilon, \delta)$ is the same regardless of the values of $\epsilon$ (the tolerable level of the excess distortion probability) and $\delta$ (the tolerable level of the overflow probability) if the sum of $\epsilon$ and $\delta$ is constant. 
That is, it holds that $R^{*} (n, D, \epsilon, \delta) = R^{*} (n, D, \epsilon', \delta')$ for $\epsilon + \delta = \epsilon' + \delta'$.
This result can be seen as the lossy version of the known result in the lossless source coding; 
in variable-length lossless source coding allowing non-vanishing error probability, it is known that the optimal rate is the same regardless of the values of $\epsilon$ (the tolerable level of the error probability) and $\delta$ (the tolerable level of the overflow probability) if the sum of $\epsilon$ and $\delta$ is constant (see, e.g., \cite[Remark 4.1]{Nomura17}).
\end{rem}

\section{Example of General Coding Theorem: Asymptotics for a Stationary Memoryless Source} \label{ASC_C}
In this section, we apply our general coding theorem for a stationary memoryless source and establish a single-letter characterization of the optimal rate $R^* (n, D, \epsilon, \delta)$.

The rate-distortion function $R_{X}(D)$ is defined by
\begin{align}
R_{X}(D) := \inf_{\substack{P_{Y |X } : \\
E[d(X,Y)] \leq D}} I(X;Y), \label{rd}
\end{align}
where $I(X;Y)$ denotes the mutual information between $X$ and $Y$.
To derive the result, the following conditions are imposed:
\begin{itemize}
\item[(i)] 
The source is stationary and memoryless.
\item[(ii)] 
For $(x^n, y^n) \in {\cal X}^n \times {\cal Y}^n$, the distortion measure $d_n(x^n, y^n)$ satisfies 
\begin{align}
d_n(x^n, y^n) = \sum_{i=1}^{n} d(x_i, y_i). 
\end{align}
\item[(iii)]
The infimum in (\ref{rd}) is achieved by $P^{\star}_{Y | X}$.
\item[(iv)]
$
E[(d (X, Y^{\star}))^{9}] < \infty,
$
where the expectation is with respect to $P_{X} \times P_{Y^{\star}}$.\footnote{
In this paper, $Y^{\star}$ is a random variable taking a value in ${\cal Y}$ whose distribution $P_{Y^{\star}}$ is the marginal of $P^{\star}_{Y | X} P_{X}$.}
\item[(v)]
$D_{\rm{min}} < \infty$,
where 
$
D_{\rm{min}} := \inf \{ D : R_{X}(D) < \infty \}. 
$
\item[(vi)]
$D \in (D_{\rm{min}}, D_{\rm{max}})$,
where $D_{\rm{max}}$ is defined as
$
D_{\rm{max}} := \min_{y \in {\cal Y}} E[d(X, y)].
$
\item[(vii)]
$\epsilon + \delta \in (0, 1)$.
\end{itemize}

A {\it $D$-tilted information} of $x \in {\cal X}$ is defined by
\begin{align}
\jmath_{X}(x,D) := \log\frac{1}{E[\exp \{ \lambda^{\star} D - \lambda^{\star} d(x, Y^{\star}) \}]},
\end{align}
where the expectation is with respect to $P_{Y^{\star}}$
and $\lambda^{\star} := - R'_{X}(D)$.
The variance of the $D$-tilted information is denoted as
$V_{X}(D)$, i.e.,
$
V_{X}(D) := {\rm Var} [\jmath_{X}(X,D)].
$
The quantity $V_{X}(D)$ is called the rate-dispersion function \cite{Kostina12}.

Now, we are ready to state the main result.
The next theorem shows the single-letter characterization of  $R^* (n, D, \epsilon, \delta)$ for a stationary memoryless source.
\begin{theorem} \label{Thiid}
Under the assumptions (i) -- (vii), we have
\begin{align}
R^* & (n, D, \epsilon, \delta) \notag \\
&=  R_{X}(D) + \sqrt{\frac{V_{X}(D)}{n}} Q^{-1} (\epsilon + \delta) + O \left( \frac{\log n}{n} \right), \label{singleletter}
\end{align}
where $Q(z) $ is $Q(z)= \int_{z}^{\infty} (1/\sqrt{2 \pi}) e^{-\frac{t^2}{2}} dt $ and $Q^{-1}(z)$ denotes its inverse function.
\end{theorem}

\begin{IEEEproof}
Let $R^{*}_{{\rm F}} (n, D, \varepsilon)$ be the minimum rate in {\it fixed-length} lossy source coding at blocklength $n$ under the condition that the excess distortion probability at distortion level $D$ is less than or equal to $\varepsilon$.
From Remark \ref{fixed}, it holds that (see also, e.g., \cite{Kontoyiannis14} and \cite{Nomura17})
\begin{align}
R^{*} (n, D, \varepsilon, 0) = R^{*}_{{\rm F}} (n, D, \varepsilon) + O \left(\frac{1}{n} \right). \label{FV}
\end{align}

Regarding $R^{*}_{{\rm F}} (n, D, \varepsilon)$, the next theorem has been established, which characterizes $R^{*}_{{\rm F}} (n, D, \varepsilon)$ by the rate-distortion function $R_{X}(D)$ and the rate-dispersion function $V_{X}(D)$.
\begin{theorem}[\cite{Kostina12}] \label{Thfixedlossy}
Under the assumptions (i) -- (vi) and the assumption $\varepsilon \in (0, 1)$, we have
\begin{align}
R^{*}_{{\rm F}} & (n, D, \varepsilon) \notag \\
& = R_{X}(D) + \sqrt{\frac{V_{X}(D)}{n}} Q^{-1} (\varepsilon) + O \left( \frac{\log n}{n} \right).
\end{align}
\end{theorem}

Now, we can derive (\ref{singleletter}) as follows:
\begin{align}
R^* & (n, D, \epsilon, \delta) \notag \\
&\overset{(a)}{=} R^* (n, D, \epsilon + \delta, 0) \\
&\overset{(b)}{=} R_{X}(D) + \sqrt{\frac{V_{X}(D)}{n}} Q^{-1} (\epsilon + \delta) + O \left( \frac{\log n}{n} \right),
\end{align}
where 
$(a)$ is due to Theorem \ref{Th_as_g} and Lemma \ref{PropertyG} (see Remark \ref{e+d}), and
$(b)$ follows from (\ref{FV}) and Theorem \ref{Thfixedlossy}.
\end{IEEEproof}

\section{Concluding Remarks} \label{Conc}
We have discussed the problem of variable-length lossy source coding under the criteria of the excess distortion probability and the overflow probability of codeword lengths.
We have derived the non-asymptotic (one-shot) and asymptotic fundamental limits of the optimal rates by using a new quantity based on the smooth max entropy.

One of the contributions of this study is the explicit code construction based on the distortion $D$-ball centered at $y$ (i.e., (\ref{BD})) in the proof of the achievability results.
It should be noted that this technique is useful for other problems.
For example, we have applied this technique to the problem of variable-length lossy source coding under the criteria of the normalized cumulant generating function of codeword lengths (see \cite{Saito18ISIT}) and the problem of guessing subject to distortion (see \cite{SaitoGuessing}).

\appendices
\section{Proof of Lemma \ref{PropertyG}} \label{PropG}
Lemma \ref{PropertyG} can be readily shown by the argument proceeded in the proof of achievability result described in Sec.\ \ref{Proof_Lem_oneshot_ac}.
More precisely, from the definition of $i^{*}$ (see (\ref{ih2}) and (\ref{ih1})), it is easy to see that $\log i^{*}$ only depends on $\epsilon + \delta$.
Combining this fact and  (\ref{iG}), we obtain the desired result.

\section{Proof of Lemma \ref{HGLemma}} \label{Append1}
To show (\ref{HG}), the following lemma is useful.
\begin{lem} \label{AppendALemma}
If $P_{Y^{\dagger}}$ which is induced by $P_{Y^{\dagger} |X}$ satisfying 
\begin{align}
\Pr \{ d(X, Y^{\dagger}) > D \} \leq \epsilon
\end{align}
majorizes any $P_{\tilde{Y}}$ which is induced by $P_{\tilde{Y} |X}$ satisfying 
\begin{align}
\Pr \{ d(X, \tilde{Y}) > D \} \leq \epsilon,
\end{align}
then it holds that
\begin{align}
H^{\delta}(Y^{\dagger}) = G_{D, \epsilon}^{\delta}(X).
\end{align}
\end{lem}

\begin{IEEEproof}
The lemma follows from the fact that the smooth max entropy is a Schur concave function
and the definition of $G_{D, \epsilon}^{\delta}(X)$.
\end{IEEEproof}

In view of Lemma \ref{AppendALemma}, we shall show that $P_{\hat{Y}}$ majorizes any $P_{\tilde{Y}}$ induced by $P_{\tilde{Y} |X}$ satisfying $\Pr \{ d(X, \tilde{Y}) > D \} \leq \epsilon$.
To show this fact, suppose the following condition:

\begin{itemize}
\item[($\spadesuit$)] {\it There exists a $P_{\tilde{Y}}$ satisfying $\Pr \{ d(X, \tilde{Y}) > D \} \leq \epsilon$ but not being majorized by $P_{\hat{Y}}$.}
\end{itemize}

Assuming ($\spadesuit$), we shall show a contradiction.

Let $y_{\pi(1)}$ give the largest $P_{\tilde{Y}}(y)$ in ${\cal Y}$, 
$y_{\pi(2)}$ give the largest $P_{\tilde{Y}}(y)$ in ${\cal Y} \setminus \{y_{\pi(1)} \}$, 
$y_{\pi(3)}$ give the largest $P_{\tilde{Y}}(y)$ in ${\cal Y} \setminus \{y_{\pi(1)}, y_{\pi(2)} \}$, etc.
That is, 
\begin{align}
P_{\tilde{Y}} (y_{\pi(1)}) \geq P_{\tilde{Y}} (y_{\pi(2)}) \geq \cdots \geq P_{\tilde{Y}} (y_{\pi(k^*)})
\end{align}
and 
\begin{align}
P_{\tilde{Y}}(y_{\pi(k^*)}) \geq P_{\tilde{Y}}(y_{\pi(i)})
\end{align}
for all $i=k^* +1, k^* +2, \ldots.$
Considering the fact that the support of $P_{\hat{Y}}$ is $\{1,2, \ldots, k^* \}$
and the assumption ($\spadesuit$), we can say that there exists a $ j_0 \in \{ 1, 2, \ldots, k^* - 1 \}$ satisfying
\begin{align}
\sum^{j_0}_{i=1} ( P_{\tilde Y} (y_{\pi (i)})  - P_{\hat{Y}} (y_i)) > 0 .\label{MU}
\end{align}

On the other hand, the excess distortion probability under 
$P_{X} P_{\tilde{Y} | X}$ is evaluated as
\begin{align}
\Pr & \{ d(X, \tilde{Y}) > D \} \notag \\
& \geq \sum_{x \in {\cal X}} \sum^{j_0}_{i=1} P_{X}(x) P_{\tilde{Y} | X}(y_{\pi (i)} | x) I \{  d(x, y_{\pi (i)}) > D \} \\
& = \sum_{x \in {\cal X}} \sum^{j_0}_{i=1} P_{X}(x) P_{\tilde{Y} | X}(y_{\pi (i)} | x)  - \sum_{x \in {\cal X}} \sum^{j_0}_{i=1} P_{X}(x) P_{\tilde{Y} | X}(y_{\pi (i)} | x)  I \{  x \in B_D (y_{\pi (i)})\} \\
&= \sum^{j_0}_{i=1}  P_{\tilde{Y}}(y_{\pi (i)} ) - \sum_{x \in {\cal X}} P_{X}(x)  \sum^{j_0}_{i=1} P_{\tilde{Y} | X}(y_{\pi (i)} | x)  I \{  x \in B_D (y_{\pi (i)})\}  \\
&  \geq \sum^{j_0}_{i=1}  P_{\tilde{Y}}(y_{\pi (i)}) - \Pr \left \{X \in \bigcup_{i=1}^{j_0} B_{D} (y_{\pi (i)}) \right \}, \label{ed}
\end{align}
where $I \{ \cdot \}$ is the indicator function and the last inequality is due to
\begin{align}
\sum^{j_0}_{i=1}  P_{\tilde{Y} | X}(y_{\pi (i)} | x) I \{  x \in B_D (y_{\pi (i)})\}  \leq I \left \{ x \in \bigcup_{i=1}^{j_0} B_{D} (y_{\pi (i)}) \right \}
\end{align}
for all $x \in {\cal X}$.
For the second term in (\ref{ed}), it holds that
\begin{align}
\Pr \left \{ X \in \bigcup_{i=1}^{j_0} B_{D} (y_{\pi (i)}) \right \} 
& \overset{(a)}{\leq}  \Pr \left \{ X \in \bigcup_{i=1}^{j_0} B_{D} (y_i) \right \} \\
& \overset{(b)}{=} \sum^{j_0}_{i=1}  \Pr \{ X \in A_D (y_i) \} \\
& \overset{(c)}{=} \sum^{j_0}_{i=1}  P_{\hat{Y}} (y_i) - \epsilon, \label{st}
\end{align}
where
$(a)$ follows from the definition of $y_i$,
$(b)$ follows from (\ref{A1}) and (\ref{A2}), and
$(c)$ follows from (\ref{y11}) and (\ref{y12}).
%the fact that
%$P_Y (y_1) = \Pr \{ X \in A_D (y_1) \} + \epsilon$ and 
%$P_Y (y_i) = \Pr \{ X \in A_D (y_i) \}$ ($i=2, \ldots, j_0$).
Plugging  (\ref{st}) into (\ref{ed}) gives
\begin{align}
\Pr \{ d(X, \tilde{Y}) > D \} 
& \geq \sum^{j_0}_{i=1} ( P_{\tilde Y} (y_{\pi (i)})  - P_{\hat{Y}}(y_i)) + \epsilon \\
&> \epsilon,
\end{align}
where the last inequality is due to (\ref{MU}).
This is a contradiction to the fact that $\Pr \{ d(X, \tilde{Y}) > D \} \leq \epsilon$.

\section{Proof of (\ref{IJK})} \label{Append2}
The first step to show (\ref{IJK}) is the following inequality:
\begin{align}
\sum^{j^{*}}_{i=1} & \Pr \{ X \in A_{D} (y_i) \} - \sum^{i^{*}}_{i=1} \Pr \{ X \in A_{D} (y_i) \}  \notag \\
&  \overset{(a)}{\leq} 1- \gamma - \delta + \Pr \{ X \in A_{D} (y_{j^*}) \}  - (1-\epsilon-\delta) \\
& = \Pr \{ X \in A_{D} (y_{j^*}) \} + \epsilon - \gamma \\
&  \overset{(b)}{\leq} \Pr \{ X \in A_{D} (y_{j^*}) \} + \Pr \{ X \in A_{D} (y_{i^* + 1}) \}, \label{jsks}
\end{align}
where $(a)$ follows from (\ref{ih1}) and (\ref{js2})
and $(b)$ follows from 
\begin{align}
\epsilon - \gamma 
& \leq \left ( 1 - \sum_{i=1}^{k^{*}-1} \Pr \{ X \in A_{D} (y_i) \} \right ) - \left ( 1 - \sum_{i=1}^{k^{*}} \Pr \{ X \in A_{D} (y_i) \} \right ) \\
& = \Pr \{ X \in A_{D} (y_{k^*}) \} \\
& \leq \Pr \{ X \in A_{D} (y_{i^* + 1}) \},
\end{align}
where the last inequality is due to $i^* < \min(j^*, k^*)$.

Inequality (\ref{jsks}) is equivalent to
\begin{align}
\sum^{j^{*}-1}_{i=1} \Pr \{ X \in A_{D} (y_i) \}  \allowbreak \leq \sum^{i^{*}+1}_{i=1} \Pr \{ X \in A_{D} (y_i) \}.
\end{align}
Thus, we obtain $j^* - 1 \leq i^* +1$, implying that 
$
\min(j^*, k^*) \leq i^* +2.
$


\begin{IEEEbiographynophoto}{Shota Saito} received the B.E.\ degree, the M.E.\ degree, and the Ph.D.\ degree in applied mathematics from Waseda University, Tokyo, Japan, in 2013, 2015, and 2018, respectively. 

From 2018, he is an Assistant Professor at the Department of Applied Mathematics, Waseda University, Tokyo, Japan.
His research interests include information theory and its applications.

Dr.\ Saito is a recipient of IEEE IT Society Japan Chapter Young Researcher Best Paper Award and Student Paper Award in the 2016 International Symposium on Information Theory and Its Applications (ISITA 2016).
He also won 2016 Waseda University Azusa Ono Memorial Award (Academic).
%(S'16--M'18)
\end{IEEEbiographynophoto}

\begin{IEEEbiographynophoto}{Hideki Yagi} received the B.E.\ degree, the M.E.\ degree, and the Ph.D.\ degree in industrial and management systems engineering from Waseda University, Tokyo, Japan in 2001, 2003, and 2005, respectively. 

He is currently an Associate Professor at the Department of Computer \& Network Engineering, The University of Electro-Communications, Tokyo, Japan. 
He was with Media Network Center, Waseda University as a Research Associate from 2005 to 2007, and an Assistant Professor from 2007 to 2008. In the winter of 2008 and from July, 2010 to January, 2011, he was a Visiting Fellow at Princeton University. 
His research interests include multi-user information and coding theory and information theoretic security.

Dr.\ Yagi is a member of the Institute of Electronics, Information and Communication Engineering (IEICE) and a member of the Research Institute of Signal Processing (RISP).
%(S'03--M'05)
\end{IEEEbiographynophoto}

\begin{IEEEbiographynophoto}{Toshiyasu Matsushima} received  the B.E.\ degree, the M.E.\ degree, and the Ph.D.\ degree in Industrial Engineering and Management from Waseda University, Tokyo, Japan, in 1978, 1980 and 1991, respectively. 

From 1980 to 1986, he joined NEC Corporation, Kanagawa, Japan. 
From 1989 to 1993, he was a Lecturer at the Department of Management Information, Yokohama College of Commerce. 
From 1993, he was an Associate Professor, and from 1996 to 2007 a Professor at the Department of Industrial and Management System Engineering, Waseda University. 
Since 2007, he has been a Professor at the Department of Applied Mathematics, Waseda University. 
From 2001 to 2002, he was a visiting researcher at the Department of Electrical Engineering, University of Hawaii, USA. 
From 2011 to 2012, he was a visiting scalar at the Department of Statistics, University of California, Berkeley, USA.
His research interests are information theory, statistics, learning theory, and their applications. 

Dr.\ Matsushima is a fellow of the Institute of Electronics, Information and Communication Engineering (IEICE) and a member of the Japan Society for Quality Control, the Japan Industrial Management Association, and the Japan Society for Artificial Intelligence.
\end{IEEEbiographynophoto}

%\newpage


%\vfill


\end{document}